# MatRec: Matrix Factorization for Highly Skewed Dataset

Hao Wang
Huahui Changtian
Beijing, China
haow85@live.com

Bing Ruan
Huahui Changtian
Beijing, China
ruanbing@tsingvm.com

## ABSTRACT
Recommender systems is one of the most successful AI technologies applied in the internet cooperations. Popular internet products such as TikTok, Amazon, and YouTube have all integrated recommender systems as their core product feature. Although recommender systems have received great success, it is well known for highly skewed datasets, engineers and researchers need to adjust their methods to tackle the specific problem to yield good results. Inability to deal with highly skewed dataset usually generates hard computational problems for big data clusters and unsatisfactory results for customers. In this paper, we propose a new algorithm solving the problem in the framework of matrix factorization. We model the data skewness factors in the theoretic modeling of the approach with easy to interpret and easy to implement formulas. We prove in experiments our method generates comparably favorite results with popular recommender system algorithms such as Learning to Rank , Alternating Least Squares and Deep Matrix Factorization.

## CCS Concepts
· Information systems~Information systems applications~Data mining

## 1. INTRODUCTION
Recommender system is overwhelmingly successful in the internet business after entrepreneurs spending many years' efforts to search for a valid business model for the technology. In the early years of the technology, it is mainly used as an integrating functionality of websites like E-commerce platforms. The technology helps companies generate higher retention rate and profits. A large group of scientists and engineers have helped to develop hundreds of recommender system models since its invention. The earliest successful models include collaborative filtering [1] and matrix factorization [2] . The stream of small modifications and micro-invention eventually leads to technological milestones like learning-to-rank models [3] , factorization machines [4] and deep learning models such as DeepFM [5].

We categorize recommender systems as shallow models and deep models. The first class incorporates shallow machine learning technologies such as matrix factorization and learning to rank , while the second class are deep learning models like Wide and Deep [6]. Although a bit of out-of-dated, shallow models are still widely used in small companies and projects where agility, usability and efficiency far outweighs boost of performance which is only economically visible for huge datasets. It is well known since the invention of the first shallow model, that data skewness and sparsity poses serious challenges for recommender system performance. The setbacks are two folds: data skewness causes problems that need special treatment in Hadoop/Spark computation; the problems also deteriorate the performance of algorithm as they usually lead to poorer results than uniformly distributed datasets.

For instance, Peter has a special interest in history books. He reads hundreds of history books ranging from *Herodotus* to *Neil Ferguson*. However, the sci-fi series *The Three Body Problem* has become extremely popular in recent years and Peter bought one of the books at the e-commerce site whose recommender system is one of his favorite book selection tools. Peter loves *The Three Body Problem* and gives it a five star rating on the website. Afterwards, he discovers the book recommender system that used to give him history book recommendations now started to give him recommendation from other genres. This is because *The Three Body Problem* is a best seller and there are so many users that read the book, their reading taste is all reflected in the user similarity computation with Peter's reading taste and spoils the recommendation result.

The data skewness problem has drawn attention from researchers years ago, and recently they started to quantify the effect of data skewness in a scientific way. Wang et. al. [7] calculates analytical formulas for data skewness and sparsity effects in both user-based and item-based collaborative filtering. Cañamares et. al. [8] models the data skewness problem in a probabilistic way. Although, quantification of the problem has just been started, the solution to this problem has a very long history. Different scientists have resorted to different heuristics and solutions to solve this issue. For example, Wang et. al. [9] points out by taking advantage of embedding vector of side information, it is possible to ameliorate the data skewness problem.

Since in the end, recommender models need to be deployable and run-time efficient in real product release environment, we seek to solve the data skewness problem in the framework of matrix factorization. We aim to solve difficult problems using simple and understandable methods that has been neglected by the community in the decade. After providing technical details in the following sections, we give experimental results of our algorithm in the end. We prove that our algorithm is not only run-time efficient, but also super ior to many well-known algorithms such as Bayesian Personalized Ranking [3] in performance.

## 2. RELATED WORK

Recommender system has a long history of evolution with many interesting and important innovations. The recommender systems have developed from shallow models such as matrix factorization [2] and learning to rank [3] to more sophisticated models that rely on deep learning theories. The main evolution theme of recommender systems is clear as optimization of evaluation metrics such as MAE, AUC and NDCG are the major driving force behind new inventions. However, major challenges such as data skewness and data sparsity remain unsolved. Techniques aiming to solve such problems usually focus on development of new models yielding better performance with no explicit or direct efforts or explanation on the particular resolution of the problems. In other words, the resolution of data skewness and sparsity problems is usually tackled indirectly without good understanding of the improvement in performance.

The first efforts to resolve the problems scientifically are done by Wang et. al. [7] and Cañamares et. al. [8] They independently model the data skewness and sparsity problems in combinatorial and probabilistic frameworks. The quantification of the problems and their effects is a major milestone in understanding the intrinsic structural problems of the recommendation algorithms. However, application of their theory remains open and challenging. With the focus shifting from evaluation metric improvement for shallow models to proposing new deep learning models in the research community, direct resolution to the problems remains largely ignored by scientists and engineers. In this paper, we propose a simple model that directly solves the data skewness and sparsity problems by integrating corresponding factors into problem modeling.

The framework selected to build in our theory is matrix factorization [2] . Although our approach can be easily extended to more sophisticated models , we choose to illustrate our ideas in a well understood and easy to implement model, which is matrix factorization. Matrix factorization is less researched in the community in recent year due to the rising interest in deep learning models, but it is still a valuable model in the industry where not every one has mastered the knowledge of deep learning and shallow models are often used as baseline models for experimental comparison.

Famous matrix factorization models include Alternating Least Squares [10] , SVD based recommendation [11] , pLSA [12] and LDA [13] . The Factorization Machines (FM) model [4] could also be considered as a generalization of matrix factorization models. The matrix factorization models could be combined with other models to yield better performance. Successful combined models like Collaborative Topic Regression [14] are deployed in commercial systems such as New York Times news recommendation. Variants of the factorization machines can be seen in many data science competition solutions and are among the most practical approaches in commercial environments.

## 3. ALGORITHM

As in social network analysis, power law distribution exists everywhere in recommender system theories and practice. The discovery of power law distribution could be traced back to Zipf's distribution. Zipf's distribution says the distribution of English words in the document corpus follows the following distribution:

A simple plot of Zipf's distribution would show that the most popular English words exhibit exponentially higher frequency than less popular English words. The distribution plot shows a highly skewed distribution with a very long tail.

The Zipf's law could be considered as the simplest power law distribution. General forms of power Law distribution exist in nearly all the input datasets to recommender systems. According to Wang et. al. [7] , power law distribution in the input data structure causes power law phenomenon in the output data structures. The resultant skewness in output could be expressed analytically as a function of the skewness in the input data structures. The variables used in their research to depict the power law distributions are the rank of popularity of the item (*item rank*) and the rank of popularity of the user (*user rank*).

Matrix factorization is a commonly adopted approach for many recommendation scenario. The basic idea is to factorize the rating matrix of users into the dot product of the user feature vector and item feature vector, namely:

$$R_{i,j} = user\ feature_i \cdot item\ feature_j$$

Different matrix factorization schemes differ in how the user feature vector and item feature vector are computed. Probabilistic models such as pLSA and Latent Dirichlet Allocation can all be applied to solve the problem. The difference in problem modeling and optimization method selection could lead to observably different performance. The paradigm is in essence a dimensionality reduction approach that reduces computation of *O(n²)* unknown variables into computation of *O(n)* unknown variables.

To borrow the concepts of Wang et. al. [7] and build the power law effect into the matrix factorization framework, the formulation of the user feature vector and item feature vector of matrix factorization is modified in the following way:

$$user\ feature_i = user\ rank_i \cdot a_i + item\ rank_j \cdot b_i + u_i$$

$$item\ feature_j = user\ rank_i \cdot c_j + item\ rank_j \cdot d_j + v_j$$

As in all the matrix factorization methods, known values in rating matrix are served as guidance for seeking unknown values of user feature vector and item feature vector, the dot product of which in turn determines the unknown values of the rating matrix. The loss function to be optimized is the sum of squared loss of rating matrix values :

$$Loss = \sum_{i=1}^{n}\sum_{j=1}^{m}(R_{i,j} - user\ feature_i \cdot item\ feature_j)^2$$

Optimization algorithms such as stochastic gradient can be applied to solve the loss function. Abiding by the common practice in the industry , we use stochastic gradient to compute the user feature vector and item feature vector as follows:

$$a = a + \eta \cdot ((2 \cdot x \cdot (R - (t_0^T \cdot t_1)/t_4))/t_4 \cdot t_1 - (2 \cdot t_5 \cdot x \cdot (R - t_5/t_4))/(t_2^3 \cdot t_3) \cdot t_0)$$

$$b = b + \eta \cdot ((2 \cdot y \cdot (R - (t_0^T \cdot t_1)/t_4))/t_4 \cdot t_1 - (2 \cdot t_5 \cdot y \cdot (R - t_5/t_4))/(t_2^3 \cdot t_3) \cdot t_0)$$

$$c = c + \eta \cdot ((2 \cdot t_6 \cdot x)/t_4 \cdot t_0 - (2 \cdot t_5 \cdot t_6 \cdot x)/(t_2 \cdot t_3^3) \cdot t_1)$$

$$d = d + \eta \cdot ((2 \cdot t_6 \cdot y)/t_4 \cdot t_0 - (2 \cdot t_5 \cdot t_6 \cdot y)/(t_2 \cdot t_3^3) \cdot t_1)$$

$$u = u + \eta \cdot ((2 \cdot (R - (t_0^T \cdot t_1)/t_4))/t_4 \cdot t_1 - (2 \cdot t_5 \cdot (R - t_5/t_4))/(t_2^3 \cdot t_3) \cdot t_0)$$

$$v = v + \eta \cdot 2 \cdot (t_6/t_4 \cdot t_0 - (t_5 \cdot t_6)/(t_2 \cdot t_3^3) \cdot t_1)$$

, where the symbols in the formulas above are defined as follows:

$$t_0 = u + x \cdot a + y \cdot b$$

$$t_1 = v + x \cdot c + y \cdot d$$

$$t_2 = \|t_0\|_2$$

$$t_3 = \|t_1\|_2$$

$$t_4 = t_2 \cdot t_3$$

$$t_5 = t_1^T \cdot t_0$$

$$t_6 = R - t_5/t_4$$

$$x = user\ rank_i$$

$$y = item\ rank_j$$

The normalization of user feature vector and item feature vector is crucial in the optimization steps of the algorithm. Failing to normalize leads easily to gradient explosion that can not be handled by computer systems.

The algorithm follows the standard matrix factorization protocol : In each iteration , a batch of or a single sample of user-item rating tuples is selected as the input dataset, then the variables of user feature vectors and item feature vectors are updated according to the formulas listed above. When prediction needs to be done, the dot product of the user feature vector and item feature vector is computed for the unknown item rating by the user.

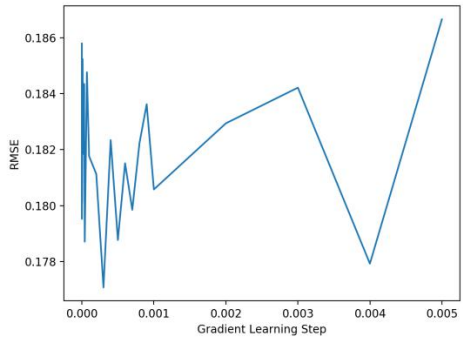

**Fig.1 MAE of the proposed algorithm for different parameters**

## 4. EXPERIMENTAL RESULTS

We set up our experiments using the lastFM [15] and MovieLens datasets . The dataset contains 1892 users and 17632 artists to be recommended to the users. The number of iterations of stochastic gradient descent is fixed to be 300, while the gradient learning step $\eta$ is enumerated to find the optimal result. The result is plotted in Fig. 1.

The lowest MAE 0.1771 is achieved when the gradient learning step is $3 \times 10^{-4}$. Bayesian Personalized Ranking (BPR) is used in our experiment for comparison with the method. The optimal MAE of BPR is 0.23609 and the MAE stays at the level of 0.2+ with varying parameters. Our algorithm is superior to the widely adopted learning to rank technique.

Fig. 2 shows the MAE of BPR algorithm with varying gradient learning steps. The number of iteration is 20 and the dimension of latent factors is 20.

Alternating Least Squares is also tested in our experiments for comparison. An MAE of 0.0518 is achieved when matrix rank is 10 and number of iterations is 10. The performance is superior to our approach , however this is only the case with the lastFM data set.

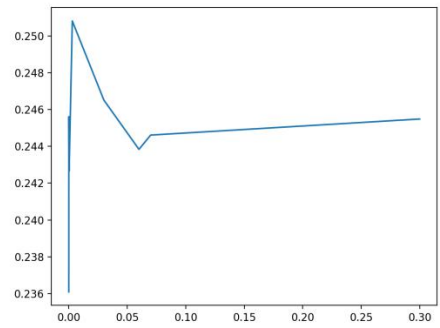

**Fig.2 MAE of Bayesian Personalized Ranking with varying gradient learning steps**

When tested on the Movie Lens dataset comprised of 162000 users and 62000 movies, a lowest MAE of 0.8618 is achieved by our method when the gradient learning step is $1 \times 10^{-4}$ (Fig. 3). The result is superior to that of Alternating Least Squares, which has an increasing MAE starting from above 0.94 when the matrix rank is 5.

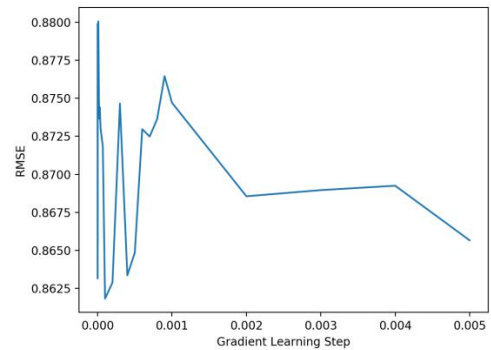

**Fig.3 MAE of proposed method in this paper**

Deep Matrix Factorization [16] generates slightly better results than our method with an MAE between 0.82 and 0.83 with selected parameters. However, the speed of Deep Matrix Factorization is much slower than our method. Without the help of GPU, it would take hours to compute the result for the same dataset on a commercial MacBook where our method takes only tens of seconds to finish. The sacrifice of speed for the slight improvement of accuracy is not always economic in real world scenarios.

## 5. CONCLUSION

In this paper, we proposed a new matrix factorization model that models the data skewness effect in input data structures explicitly. The algorithm is invented in hope to ameliorate the famous data skewness problem for recommender system. In our experiments, we resort to conventional evaluation metrics for algorithm evaluation. Our method has comparatively favorite result in contrast with Learning to Rank approaches such as BPR and matrix factorization methods like Alternating Least Squares and Deep Matrix Factorization. However, the metric we used in experiments is MAE, which might not be the most appropriate metric to evaluate recommender system performance in consideration of data skewness. In future work, we would like to create theoretical foundation to solve the data skewness problem for recommender system including finding better evaluation metrics for experiments.